\documentclass{aa}  
\usepackage{natbib}
\usepackage{xspace}
\usepackage{amssymb}
\usepackage{amsmath}
\usepackage{graphicx}
\usepackage{comment}

\def\aap{A\& A}

\def\mnras{MNRAS}

\def\apj{ApJ}

\def\ffric{\ensuremath{F_\mathrm{fric}}\xspace}

\def\cs{\ensuremath{c_\mathrm{s}}\xspace}

\def\rzero{\ensuremath{r_0}\xspace}
\def\Omegak{\ensuremath{\Omega_\mathrm{K}}\xspace}

\def\Sc{\ensuremath{\mathrm{Sc}}\xspace}

\def\vacc{\ensuremath{v_\mathrm{acc}}\xspace}
\def\tdiff{\ensuremath{t_\mathrm{diff}}\xspace}
\def\tacc{\ensuremath{t_\mathrm{acc}}\xspace}

\def\Lstar{\ensuremath{L_\star\xspace}\xspace}

\def\vth{\ensuremath{v_\mathrm{th}\xspace}\xspace}
\def\sigpile{\ensuremath{\sigma_\mathrm{pile}\xspace}\xspace}
\def\rpile{\ensuremath{r_\mathrm{pile}\xspace}\xspace}
\def\ffric{\ensuremath{f_\mathrm{fric}\xspace}\xspace}
\def\Ffric{\ensuremath{F_\mathrm{fric}\xspace}\xspace}
\def\frad{\ensuremath{f_\mathrm{rad}\xspace}\xspace}
\def\Frad{\ensuremath{F_\mathrm{rad}\xspace}\xspace}
\def\vdrift{\ensuremath{v_\mathrm{dr}\xspace}\xspace}
\def\fdg{\ensuremath{f_\mathrm{dg}\xspace}\xspace}
\def\Mdot{\ensuremath{\dot{M}\xspace}\xspace}
\def\rstop{\ensuremath{r_\mathrm{stop}}\xspace}
\def\tstop{\ensuremath{t_\mathrm{stop}}\xspace}

\begin{document}
%\thesaurus{02.01.2,08.03.4,08.06.2,08.16.5,13.09.6}
%\title{Can residual infall replenish the small particle population in a disk?}
\title{Accretion through the inner hole of transitional disks:\linebreak
  What happens to the dust?}
\titlerunning{Accretion through dist holes: What happens to the dust?}
\authorrunning{Dominik, Dullemond}
\author{C.~Dominik$^{1,2}$ \& C.P.~Dullemond$^{3,4}$}
\institute{$^{1}$Sterrenkundig Instituut `Anton Pannekoek', Science
  Park 904, NL-1098 XH Amsterdam, The Netherlands; e--mail:
dominik@uva.nl\\
$^{2}$Afdeling Sterrenkunde, Radboud Universiteit Nijmegen,
Postbus 9010, 6500 GL Nijmegen\\
$^3$Institut f\"ur Theoretische Astrophysik, Universit\"at Heidelberg
Albert-Ueberle-Str. 2, D-69120 Heidelberg, Germany; e--mail: 
dullemond@uni-heidelberg.de\\
$^{4}$Max Planck Institut f\"ur Astronomie, K\"onigstuhl 17,
D-69117 Heidelberg, Germany}
\date{DRAFT, \today} 

\abstract{We study the effect of radiation pressure on the dust in the
  inner rim of transitional disks with large inner holes.  In
  particular, we evaluate whether radiation pressure can be
  responsible for keeping the inner holes dust-free, while allowing
  gas accretion to proceed.  This has been proposed in a paper by
  Chiang and Murray-Clay (2007, Nature Physics 3, p. 604) who explain
  the formation of these holes as an inside-out evacuation due to
  X-ray-triggered accretion of the innermost layer of the disk rim
  outside of the hole.  We show that radiation pressure is clearly
  incapable of stopping dust from flowing into the hole because of
  dust pile-up and optical depth effects, and also because of viscous
  mixing. Other mechanisms need to be found to explain the persistence
  of the opacity hole in the presence of accretion, and we speculate
  on possible solutions.}

\maketitle


\section{Introduction}
\label{sec:introduction}
In recent years our understanding of the structure of the dusty disks
around young stars has increased considerably with the enormous data
volumes of such disks from space-based infrared telescopes (ISO,
Spitzer Space Telescope), ground-based infrared interferometry, and
millimeter wave observations.  One interesting issue that emerged is
that a certain fraction of these protoplanetary disks appear
to have large inner holes.  To be more precise, these objects have
inner disks that are nearly dust-free out to several AU, much farther
out than the dust sublimation radius.  
Such disks are frequently termed ``transitional disks'', expressing
the idea that these disks are in the process of becoming debris disks.

Several ideas for the origin of these holes are being discussed.  One
idea is that photoevaporation might have drilled a hole in the inner
disk
\citep{2005prpl.conf.8433H,2006MNRAS.369..216A,2007MNRAS.376.1350C,2008ApJ...688..398E}.
Another explanation is the presence of planetary systems inside the
``gap'' \citep[e.g.][]{2009ApJ...704..989A}.

Surprisingly, several transitional disks still show signs of accretion
\citep[e.g.][]{2009A&A...504..461F,2010ApJ...710..597S}, indicating
that some gas is streaming through these opacity holes without the
usual dust content.  Chiang \& Murray-Clay
(\citeyear{2007NatPh...3..604C}, henceforth CM07) propose a mechanism
by which the ionizing effect of X-rays from the star causes an
inside-out evacuation of the disk.  In this scenario, radiation
pressure on dust grains plays a crucial role in holding back the dust
while allowing the gas to flow in.  CM07 make a detailed analysis of
the conditions needed to activate the magneto-rotational instability
(MRI) necessary for accretion.  In their model, the disk is largely
viscously inactive because of insufficient ionization.  Only directly
at the inner rim of the disk, X-rays at energies of about 3\,keV from
the central star lead to sufficient ionization in a thin layer with
column density $\sim 5\times10^{23}$cm$^{-2}$.  As this layer
viscously spreads inward and is being accreted by the star, new layers
get exposed to the ionizing radiation.  In this way, the disks gets
evacuated from the inside out, leading to a steadily growing inner
hole.  CM07 find accretion rates that range from 10$^{-11}$ to
10$^{-8}$\,M$_{\odot}$/yr.  If the dust was dragged along with the
gas, and if the accretion onto the star occurs in a continuous way
through a disk, then there will be no opacity hole.  CM07 propose that
radiation pressure on dust grains is the mechanism to keep the inner
disk free of dust.  While CM07 consider the possibility that some dust
leaks into the inner hole, there is no mechanism to eventually remove
it again from the hole.

Here we revisit this issue of holding back the dust by
radiation pressure.  Our main concern is the optical depth effect,
which limits the total amount of dust that can be affected directly by
radiation from the central star.  As the accretion layer proceeds into
the disk, dust will accumulate and eventually collectively overcome
the radiation pressure.

\section{Model}

Let us consider a radial column of gas and dust in the inner disk rim,
exposed to radiation from the star.  Radiation pressure acts
only on dust grains in the inner few optical mean free paths
($\tau_{\star}\lesssim\mbox{few}$), with the radiation force
decreasing with increasing optical depth.  Ignoring other effects, the
innermost grains would therefore quickly catch up with particles at
higher optical depth, creating an infinitely thin sheet of dust.
Particle diffusion (either thermal or turbulent) will act to keep the
thickness of this sheet finite.  Furthermore, it will also transport
the effect of radiation pressure to the dust deeper in the disk,
thereby allowing a larger amount of dust to be affected. The radiation
pressure will thus act as an outward moving ``snow plow'', piling up
more and more dust.

As the dust particles are pushed outward through the gas, they will
feel a friction force.  The total friction force on the pile will
increase as the pile builds up. On the other hand, the
radiation pressure is fixed.  All particles in the pile will have
to share this given amount of radiation pressure.  The dust pile
will therefore gradually slow down.  Even a very small
inward pointing accretion flow will eventually stop the snow plow
and then drag the dust pile inward with the gas.

\subsection{Analytical estimates}
\label{sec:analytical-estimates}

\begin{comment}
%  \subsection{Basic equations for disk accretion}

  \begin{equation}
    \label{eq:1}
    \Sigma = - \frac{\dot{M}}{2\pi R \vacc} = \frac{\dot{M}}{3\pi\nu}
  \end{equation}
  \begin{equation}
    \label{eq:2}
    \vacc = -\frac{3}{2}\frac{\nu}{R}
  \end{equation}
  \begin{equation}
    \label{eq:3}
    \dot{M}= 2\pi R \Sigma \vacc
  \end{equation}
  \begin{equation}
    \label{eq:4}
    D=\frac{\nu}{\Sc}
  \end{equation}
  \begin{equation}
    \label{eq:5}
    \nu = \alpha \frac{\cs^2}{\Omegak}
  \end{equation}
\end{comment}

\subsubsection{When does a pile-up take place at all}

The radiation force of the unattenuated stellar luminosity $\Lstar$
acting on a single grain with radius $a$ in the rim surface located at
a distance $\rzero$ from the star is given by $\Frad =
L_{\star}Q_{\mathrm{pr}}\pi a^2/(4\pi \rzero^2c)$ where
$Q_{\mathrm{pr}}$ is the momentum absorption efficiency of the grain
and $c$ is the speed of light.  For the analytical and numerical
estimates below, we take $Q_{\mathrm{pr}}=2$, to simulate the most
favorable condition for holding grains back with radiation pressure.
At the same time, the gas is streaming slowly toward the star, with
the accretion velocity $\vacc$. In the Epstein regime and for
  subsonic dust drift, we can write the friction force per grain as
  $\Ffric = 4\vacc\vth\rho\pi a^2/3$ \citep[e.g.][]{Schaaf63} where
$\rho$ is the gas density and $\vth=\sqrt{8 kT/(\pi \mu m_p)}$ is the
thermal velocity of gas particles with mass $\mu m_p$ at a temperature
$T$.  $m_p$ is the proton mass, and we take $\mu=2.3$ for molecular
gas.

The first question we need to answer is whether there will be any
pile-up at all.  This is the case if the radiation force onto a
single grain exceeds the friction force onto that same grain.  The
criterion for this is
\begin{equation}
\label{eq:8}
\gamma := \frac{3}{16\pi}\frac{\Lstar
  Q_{\mathrm{pr}}}{\rzero^2c\vacc\vth\rho} > 1
\quad .
\end{equation}
If $\gamma<1$, the dust will accrete with the gas onto the star
without delay; i.e., the space velocity of the dust grains will never
point away from the star.  If $\gamma>1$, on the other hand, the
accretion of dust will be delayed until a large enough pile has
accumulated.

\subsubsection{Speed of the snow plow relative to the gas}

For the moment we will assume that the gas in the outer disk is not
moving at all, and that the radiation pressure is simply driving the
snow plow a distance $s$ into the disk, starting at a position \rzero.
The speed of the column piled up by radiation pressure can be
calculated by equating the radiation force onto a column with 1 cm$^2$
inner surface $\frad = L_{\star}/(4\pi \rzero^2 c)$, with the
inward-directed collective friction force $\ffric$ of all the grains in the pile while the pile is
moving relative to the gas with a velocity \vdrift.  The latter force
is  $\ffric =\sigpile \vdrift \vth \rho/(\xi a)$,
where \sigpile is the column mass density of dust in the pile and
$\xi$ is the specific density of the grains which we take to be
2\,g/cm$^2$.  Solving for the drift velocity we find
\begin{equation}
\label{eq:11}
\vdrift = \frac{\Lstar\xi a}{4 \pi \rzero^2 c \sigpile \vth \rho}
\quad .
\end{equation}
\subsubsection{Time-dependent solution and total mass supported by
  radiation pressure}
For simplicity we assume that the gas density $\rho$ and the
dust-to-gas ration $\fdg$ are constant prior to buildup of the pile.
We define a variable $s$ that measures the distance from the rim into
the disk, so that we can write the position of the dust pile as
$\rpile(t)=r_0+s(t)$.  Then the dust mass in the pile can be expressed
easily as $\sigpile(s) = \rho \fdg s$.  The drift velocity at location
$s$ into the disk is then
\begin{equation}
\label{eq:13}
\vdrift(s) = \frac{\Lstar \xi a}{4 \pi \rzero^2 c \rho^2 \fdg \vth s}
 =: \frac{K}{s}
\quad ,
\end{equation}
where $K:=\Lstar\xi a/(4 \pi \rzero^2 c \rho^2 \fdg \vth)$.  As we can see,
the velocity of the pile becomes lower and lower as it collects
more dust to be pushed forward.
We can integrate \eqref{eq:13} to find the distance traveled by the
pile to be $s(t) = \sqrt{2Kt}$.

Now we introduce a constant inward gas motion with velocity \vacc.
The position of the pile now becomes $\rpile(t)=r_0+s(t)-\vacc t$,
while $s(t)$ remains unchanged.  The snowplow comes to a halt at
$\tstop = K/(2\vacc^2)$ and 
$\rstop = r_0+K/\vacc-\vacc \tstop$.
After that, continued mass loading will turn around the motion of the
pile and move it, embedded into the accretion flow, into the inner
hole and toward the star.  The total column at the turn-around point
is given by
\begin{equation}
\label{eq:24}
\sigpile(s_{\mathrm{stop}}) = \frac{\rho \fdg K}{\vacc} =
\frac{\Lstar\xi a}{4 \pi \rzero^2 c \rho \vth \vacc}
\quad .
\end{equation}
\begin{table*}
\caption{\label{tab:numbers}Some numerical estimates for observed
  transitional disks}
\begin{center}
% BEGIN RECEIVE ORGTBL numbers
\begin{tabular}{lrrrrrrrrr}
Star & $M_{\star}$ & $\Lstar$ & $\Mdot$ & $\alpha$ & $r_0$ & $\gamma$ & $\sigpile$ & $\rstop/r_{0}$ & $\tstop$ \\
 & [$M_{\odot}$] & [$L_{\odot}$] & [$10^{-10}M_{\odot}$/yr] &  & [AU] &  & [$10^{-5}$g/cm$^{2}$] &  & [yr] \\
\hline
DM Tau & 0.47$^{(1)}$ & 0.34$^{(2)}$ & 10.0$^{(1)}$ & 0.035$^{(1)}$ & 3$^{(1)}$ & 0.69 & 1.4 & 1.00008 & 0.6 \\
TW Hya & 0.6$^{(1)}$ & 0.23$^{(3)}$ & 4.0$^{(1)}$ & 0.01$^{(1)}$ & 4$^{(1)}$ & 1.19 & 2.4 & 1.0001 & 4.3 \\
GM Aur & 1.2$^{(1)}$ & 1.33$^{(2)}$ & 50.0$^{(1)}$ & 0.007$^{(1)}$ & 25$^{(1)}$ & 0.97 & 1.9 & 1.00004 & 15.3 \\
HD100546 & 2.5$^{(4)}$ & 36$^{(4)}$ & 10.0$^{(4)}$ & 0.035$^{(5)}$ & 10$^{(4)}$ & 57.57 & 115.1 & 1.03464 & 602.4 \\
\end{tabular}
% END RECEIVE ORGTBL numbers
\end{center}
References: (1) CM07; (2) \citet{2010A&A...512A..15R}; (3) \citet{2002ApJ...568.1008C}; (4)
\citet{2003A&A...398..607D}; (5) copied from DM Tau
\end{table*}
\begin{comment}
#+ORGTBL: SEND numbers orgtbl-to-latex :splice nil :skip 0
| Star     |   $M_{\star}$ |      $\Lstar$ |                  $\Mdot$ | $\alpha$ | $r_0$ | $\gamma$ |            $\sigpile$ | $\rstop/r_{0}$ | $\tstop$ |
|          | [$M_{\odot}$] | [$L_{\odot}$] | [$10^{-10}M_{\odot}$/yr] |          |  [AU] |          | [$10^{-5}$g/cm$^{2}$] |                |     [yr] |
|----------+---------------+---------------+--------------------------+----------+-------+----------+-----------------------+----------------+----------|
| DM Tau   |          0.47 |          0.34 |                     10.0 |    0.035 |     3 |     0.69 |                   1.4 |        1.00008 |      0.6 |
| TW Hya   |           0.6 |          0.23 |                      4.0 |     0.01 |     4 |     1.19 |                   2.4 |         1.0001 |      4.3 |
| GM Aur   |           1.2 |          1.33 |                     50.0 |    0.007 |    25 |     0.97 |                   1.9 |        1.00004 |     15.3 |
| HD100546 |           2.5 |            36 |                     10.0 |    0.035 |    10 |    57.57 |                 115.1 |        1.03464 |    602.4 |
\end{comment}

\subsubsection{Width of the peak from diffusion}
We can also estimate the width $\Delta R$ of the dust pile as it is
produced by the interplay between accretion and viscosity.  The
diffusion time across $\Delta R$ is given by $\tdiff = \Delta R^2/D$,
and the accretion time across the same distance is $\tacc = \Delta
R/\vacc$.  Equating these timescales, and using $D=\nu/\Sc$ where $D$
is the diffusion coefficient, $\nu$ is the viscosity (molecular or
turbulent) and Sc is the Schmidt number, we find $\Delta R =
\nu/(\Sc\,\vacc)$.  Please see below for more details on the Schmidt
number we use for our calculations.

\subsection{Numerical values}

Table~\ref{tab:numbers} shows numerical values for the quantities
derived above.  To derive these values we have relied on standard
accretion theory \citep[e.g.][]{2009apsf.book.....H} and used the
following relations: $\vacc=3\nu/(2r)$, $\Sigma = \dot{M}/(2\pi R
\vacc)$, and $\rho=\Sigma/(\sqrt{2\pi}\times h_p)$ with the pressure scale
height $h_p:=c_s/\Omegak$.

The table shows $\gamma$, $\sigpile$, $\rstop$, and $\tstop$ for the
three stars in CM07 and also for HD\,100546, a Herbig star that also
shows a large gap even though is does have some hot dust close to the
star \citep{2010A&A...511A..75B}.  We have included this star because
of its higher luminosity, which is helpful for illustrating the
pile-up effect.  The value of $\gamma$ shows that the radiation force
on individual grains does \emph{not greatly exceed} the friction force
created by the accretion flow - something one would have expected to
achieve significant pile-up.  Consequently, the total column of dust
piled up before the turn-around point is only of the order of
$1\times10^{-5}$ g cm$^{-2}$, and the location where this happens is
no more than a few times 10$^{-4}$ AU, very close indeed to the
original starting point of the rim.  The pile-up happen on a time
scale between 1 and 40 years.  So even starting with a clean inner
disk, dust will start to flow into the gap after that time.

The higher luminosity of the Herbig star HD\,100546 does have a
slightly stronger effect, but it will be not any more noticeable than in the
case of T Tauri stars.

\section{Numerical model}
\label{sec:nummodel}

\subsection{Model description}
The calculations so far are based on a number of simplifying
assumptions to make the problem analytically tractable.  We also
constructed a numerical model to study the radial dust motion in this
situation.  Instead of solving the complete coupled set of
equations involving both the dust {\em and} the gas motion, our goal
is less ambitious.  We want to see whether the radiation pressure can
counter the dust pile-up enough to keep the inner hole
dust-free.  Therefore, we study the reduced problem of radial 1-D dust
motion with radiation pressure in a disk with a given inward gas
motion.

The model is built in the following way.  We start by setting up the
gas density $\rho_g(r)$ at the disk midplane as a function of the
radial coordinate $r$.  The gas density is computed
semi-selfconsistently by specifying the accretion rate $\dot M$ in the
disk, the turbulence parameter for accretion $\alpha_a$ and the
flaring angle $\varphi$ for stellar irradiation.  Using the flaring
angle, the midplane disk temperature is estimated as
\citep{1997ApJ...490..368C}: $T_\mathrm{mid}(r)=
\varphi^{1/4}(r_{*}/r)^{1/2}T_{*}$, where $R_{*}$ is the stellar
radius and $T_{*}$ is the stellar effective temperature.  We assume
that no viscous heating takes place (valid for low accretion rates and
large inner holes) and that the full $\pi R_{*}^2$ stellar surface is
``visible'' at any point on the disk surface.  From the resulting
temperature we can now compute the isothermal sound speed
$c_s=\sqrt{kT/\mu m_p}$, where $k$ is the Boltzmann constant and
$\mu\simeq 2.3$ is the mean molecular weight of the gas and $m_p$ is
the proton mass.  The accretion viscosity coefficient is now given as
$\nu_a=\alpha_a c_s^2/\Omega_K$, where $\Omega_K=\sqrt{GM_{*}/r^3}$
with $M_{*}$ the stellar mass. Finally, with $\dot
M=3\pi\Sigma_g\nu_a$ (valid for $r\gg R_{*}$) the gas surface density
$\Sigma_g$ can be computed for given value of the accretion rate $\dot
M$.  For more details on standard accretion disk theory, we refer to
the book by Hartmann (\citeyear{2009apsf.book.....H}).  We do included
neither the effect of the dust motion nor of changing dust-to-gas
ratios, in our estimation of the gas disk structure.

\begin{figure}
\centerline{\includegraphics[width=8cm]{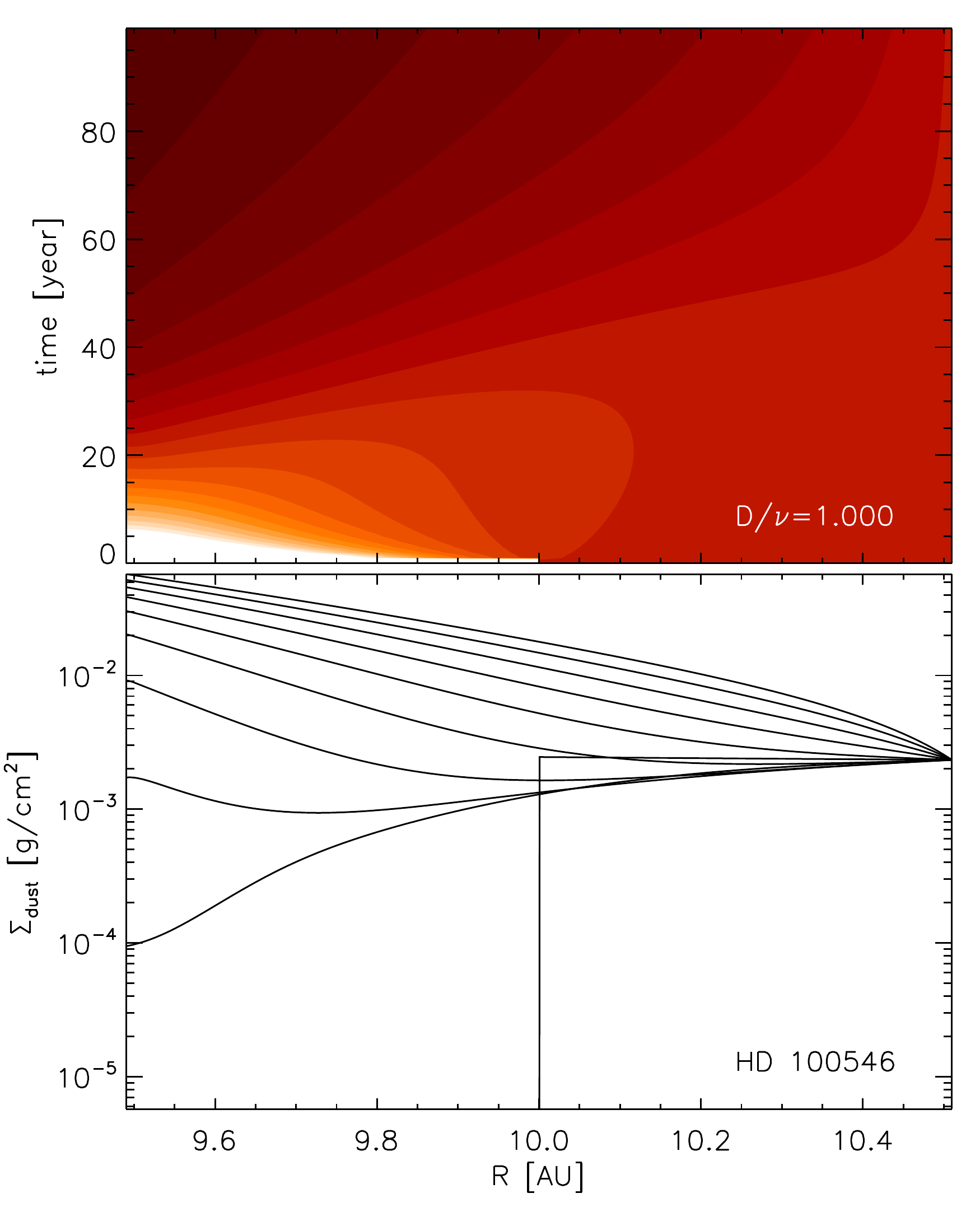}}
\caption{\label{fig-hd100546-sc1}Results of the numerical model for the
  source HD100546 for the case Sc=1, i.e.\ $D_t=\nu_t$ and
  $\alpha_t=0.035$. Top panel: Surface density of dust as a function of
  radius (horizontal axis) and time (vertical axis). The color is the dust
  density in normalized units.  The contour levels of the diagram are
  steps of 10$^{0.2}$, i.e. five contours are one factor of 10.  The level
  in the bottom-right corner is the initial value of the dust surface
  density, $2.3 \times 10^{-3}$ g/cm$^2$.
  The bottom panel is the same, but now plotted as different curves for
  different times, with the vertical axis being the dust surface
  density.  The model actually calculates the density of the dust
  $\rho_d$, but for the plotting we converted it to surface density
  $\Sigma_d=\sqrt{2\pi}H_p\rho_d$. In this model the curves that are
  more to the top are the later ones. The time interval between two
  curves is 10 years.}
\end{figure}

\begin{figure}
\centerline{\includegraphics[width=8cm]{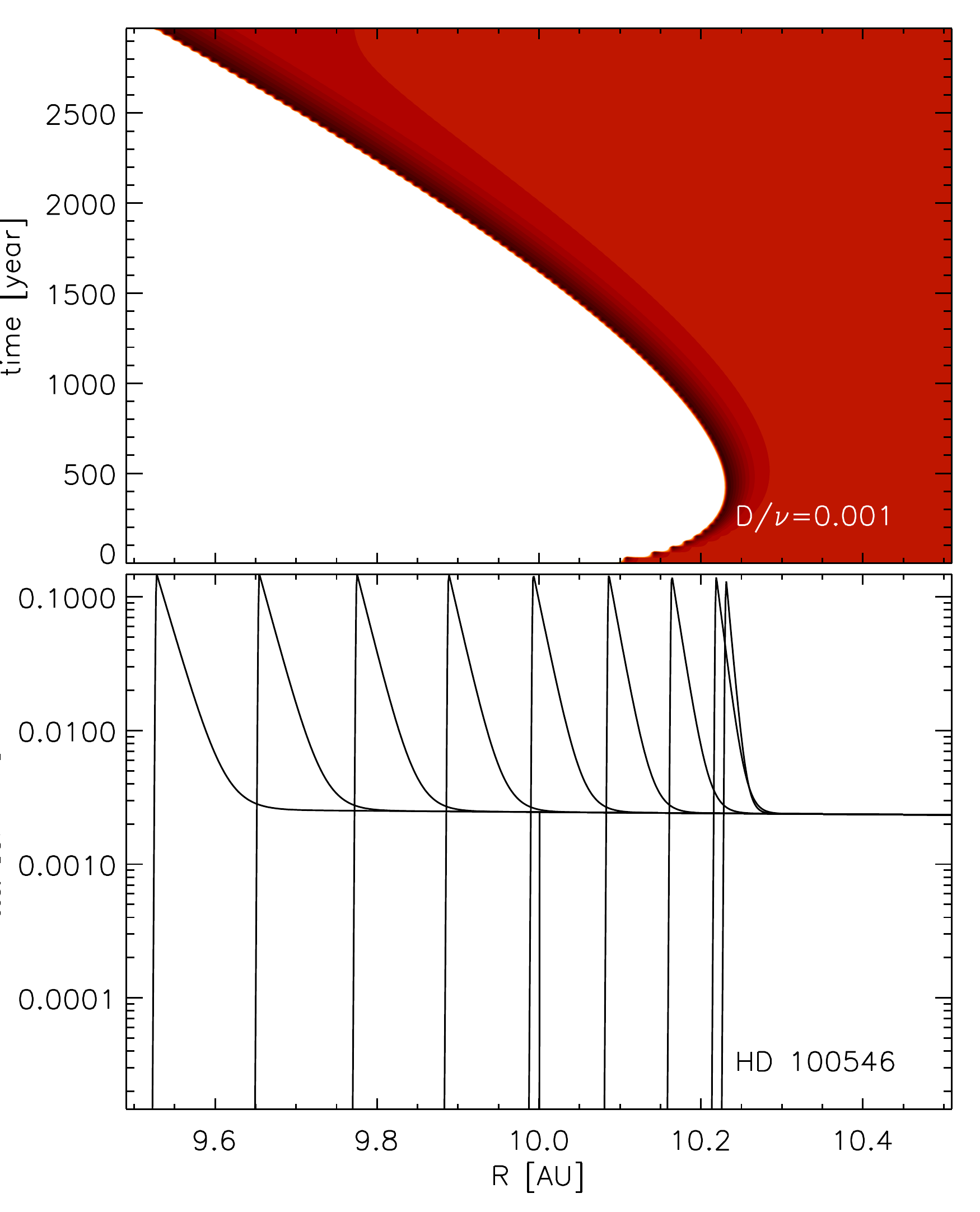}}
\caption{\label{fig-hd100546}As Fig.~\ref{fig-hd100546-sc1}, but now for
  Sc=$10^3$, i.e.\ $D_t=0.001\,\nu_t$. Time differences between curves is
  300 years this time.}
\end{figure}
We then insert the dust into this model. We assume that the dust
grains are small enough that they do not settle much, so that the
increase in midplane dust density due to settling is negligible. We
focus solely on the radial motion of the dust. At each time step we
start with a very simple radiative transfer calculation: We follow the
radiation of the star as it moves along the disk equatorial plane
outward, i.e.\ we integrate the equation$
\frac{1}{r^2}\frac{d(r^2L(r))}{dr} = -\rho_d(r,t)\kappa_d L(r)$, where
$\rho_d(r,t)$ is the dust density as a function of spatial position
$r$ and time $t$.  $\kappa_d$ is the Planck-mean dust opacity,
computed with the stellar temperature as reference temperature.  For
simplicity we assume that $\kappa_d$ is independent of $r$, i.e.\ we
neglect the changes in stellar spectrum as it is extincted in the
disk.  In fact, we do not solve for the frequency-dependent luminosity
$L_\nu(r,t)$, but instead for the frequency-integrated one
$L(r,t)$. This exerts a radiation pressure force on the dust grains,
in units of dyne/gram-of-dust, of $f_r(r,t)=\kappa_dL(r,t)/(4\pi c
r^2)$.  The dust has friction with the gas and we compute the dust
velocity by assuming that radiative and friction forces are in
equilibrium, i.e. that the dust drifts relative to the accreting gas
at its equilibrium drift velocity.

In addition to this systematic dust motion, there is also random
motion due to the gas turbulence.  While turbulence may be extremely
weak in a dead zone, we shall see that even the slightest turbulence
will profoundly affect the structure of the dust pile-up region.  The
turbulent ``diffusion'' of dust is due to the same turbulence that is
believed to be responsible for the accretion process (the turbulent
``viscosity'' $\nu_a$, see above). The turbulent diffusion constant
$D_t$ is therefore of the same order of magnitude as $\nu_a$. Exactly
what the ratio $D_a/\nu_a$ is is not perfectly understood yet
\citep{2005ApJ...634.1353J}.  If viscosity is due to turbulent gas
motions, this ratio should be not far from unity.  However, it may
also be that large scale magnetic fields could remove angular momentum
from the disk without inducing much turbulent mixing
\citep{2008A&A...479..481C}.  Here we parametrize it as the Schmidt
number: $D_t = \nu_t/\mathrm{Sc}$.  We treat Sc as a free tuning
parameter.

The dust motion now follows from the continuity equation:
\begin{equation}
\begin{split}
\frac{\partial \rho_d(r,t)}{\partial t}
&+\frac{1}{r^2}\frac{\partial(r^2\rho_d(r,t)v_d(r,t))}{\partial r}\\
&-\frac{1}{r^2}\frac{\partial}{\partial r}\left(D_tr^2\rho_g(r)
\frac{\partial}{\partial r}\left(\frac{\rho_d(r,t)}{\rho_g(r)}\right)\right)=0
\end{split}
\end{equation}
which we solve using implicit integration (``backward Euler'').

As an initial condition for the dust we set $\rho_d$ to $0.01\;\rho_g$
for $r\ge r_{0}$ and $\rho_d=0$ for $r< r_{0}$, where $r_0$ is the
initial dust inner rim radius, given in table \ref{tab:numbers}.  We
take the dust grains to be $a=0.1\,\mu$m in radius, consisting of
material with a density of $\xi=3$ g cm$^{-3}$. For simplicity we take
a simple geometric opacity of $\kappa=3/(\xi a)$, which is twice the
geometric cross section.  As boundary condition at the inner edge of
the computational domain we take a simple outflow condition, while at
the outer edge of the domain we set $\rho_d=0.01\rho_g$ as a boundary
condition.

\begin{figure}
\centerline{\includegraphics[width=8cm]{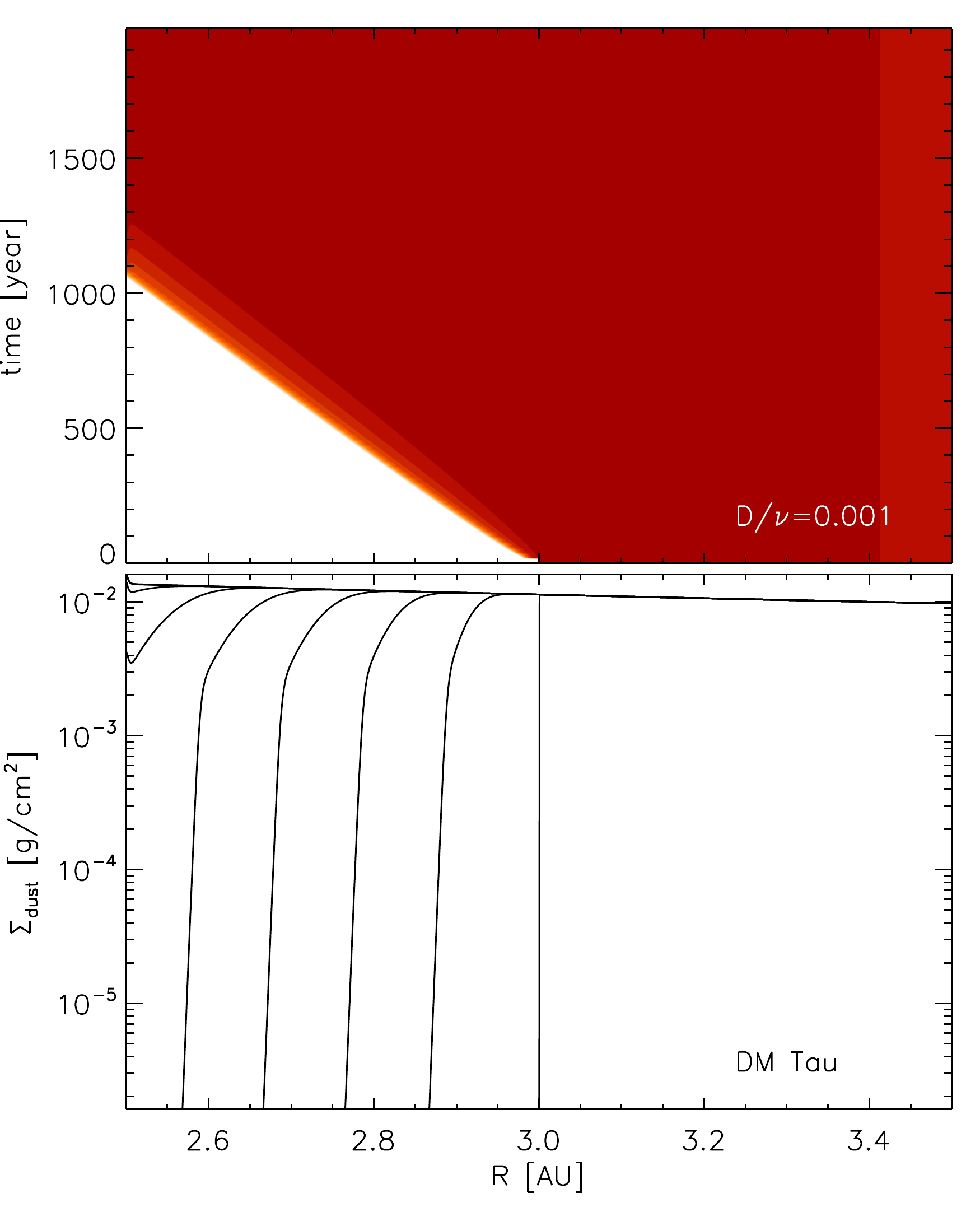}}
\caption{\label{fig-dmtau}As Fig.~\ref{fig-hd100546}, but now for the
  source DM Tau. Time interval between curves is 200 years.  The
  contour levels of the top panel are steps of $10^{0.2}$, i.e.\ five
  contours are one factor of 10.  The level in the bottom-right corner
  is the initial value of the dust surface density, $1\times10^{-2}$
  g/cm$^2$.}
\end{figure}

\subsection{Model results}
We performed the numerical equivalent models for all sources listed in
table \ref{tab:numbers}.  One of the most striking results, found in
all models, is that if we take Sc=1, i.e.\ $D_t=\nu_t$, then any
outward movement of the dust induced by the radiative pressure is
completely overwhelmed by the enormously strong turbulent diffusive
mixing.  So even for cases where the radiative force would easily push
out the grains if no turbulent mixing would occur
(Sc$\rightarrow\infty$), the grains are washed inward as a result of
strong mixing.  An example of this is shown in
Fig.~\ref{fig-hd100546-sc1}.  In this example the time scale for mixing
dust toward the inner edge of the computational domain at 9.5 AU is
about 10 years only.  The overall solution quickly becomes dominated
by the outer boundary condition (best seen in the bottom panel),
rendering it meaningless at that point.

Why is particle mixing so dominant, almost independent of $\alpha_t$?
This can be seen in the following way: If, we take strong/weak
turbulence, e.g.\ $\alpha=0.1$ / $\alpha=10^{-6}$, then, for a given
accretion rate $\dot M$, the gas surface density becomes low/high (by
virtue of $\dot M=3\pi\Sigma\nu_t$).  So for a given radiation
pressure force on the grains, the drift speed is high/low, but also
the turbulent mixing as well as the inward gas velocity become
high/low.  Working out the numbers for the accretion rates given in
table \ref{tab:numbers}, in all cases the turbulent diffusion wins
over radiation pressure.  The conclusion is that with
$\mathrm{Sc}\simeq 1$, the radiation pressure is entirely unable to
prevent the dust from drifting inward toward the star.

However, as discussed earlier, if angular momentum transport is not
primarily caused by turbulence, much larger Schmidt numbers are
allowed.  We choose Sc=$10^{3}$, but keep the $\alpha_t$ the same as
before.  Then we find that for some cases, notably HD\,100546, the
radiation pressure is strong enough to push the dust outward against
both the accretion flow of the gas and the action of turbulent
diffusion.  For other cases the gas drag of the inward gas motion is
\emph{still} too strong, such as for DM Tau. The results for
HD\,100546 and DM Tau are shown in Figs.(\ref{fig-hd100546},
\ref{fig-dmtau}).

For HD\,100546 one can see that the radiation pressure pushes the grains that
are in plain sight of the star outward. But the absorption of the radiation
(the same process that transfers radiative momentum to the grains) also
shields the dust grains at larger optical depth from stellar light.  These
dust grains therefore still flow inward with the gas. What happens is, as
one can see in the figure, that the dust piles up in a rim of high dust
density. According to our model, if one takes an even higher Schmidt number
this dust pile-up rim becomes geometrically thinner. However, the total mass
in this rim (and thus its optical depth) is only dependent on the amount of
dust that has been swept up. The radiation force acts only on the innermost
layer of this dust pile-up rim, i.e.\ the layer of radial optical depth
unity. But the diffusion, even if ever so weak, will distributed the
momentum that is received by these inner dust grains to the entire mass of
dust in the pile-up rim. One also clearly sees that the dust pile-up rim
first moves outward, but then slows down as more and more of the radiation
momentum has to be divided over an ever increasing mass of dust in the
pile-up rim. This dilution of the radiation force means that the outward
motion of the rim stalls, and eventually the dust gets dragged inward with
the gas flow.

For DM Tau we see that, even with a high Schmidt number, a pile-up
never really happens, because the gas drag by the inward moving gas is
simply too strong. One sees a nearly instant inward motion taking
place, as expected based on the $\gamma$ value in Table
\ref{tab:numbers} being lower than unity.

\section{Discussion and conclusions}
\label{sec:discussion}

In summary, we find strong arguments that radiation pressure alone is
wholly insufficient to keep the inner hole of a transitional disk
dust-free while low-level accretion is continuing.  Both dust pile-up
and viscous mixing are easily able to overcome the radiation pressure
quickly and refill the inner hole with dust on a level similar to
normal dust-to-gas ratios.  These processes pose a problem for the
inside-out evacuation scenario for transitional disk in its original
form.  While we do think the MRI activation discussed by
\citet{2007NatPh...3..604C} may be responsible for launching a
low-level accretion from the inner rim, mechanisms other than
radiative forces have to be invoked to keep dust from flowing into the
inner holes, or at least from contributing to the opacity in these
holes.

An additional force to hold back dust that has been discussed in the
literature is photophoresis \citep{2005ApJ...630.1088K}, which is a
strong force caused by the temperature difference between the
illuminated and dark sides of a dust grain.  This force may have an
effect for larger grains (cm-sizes), but it is irrelevant for the
sub-micron-sized grains considered here (see figure 1 from
\citet{2005ApJ...630.1088K}).

One possible solution is that the dust has grown to such large sizes
in these regions that the dust continuum optical depth drops below
unity\citep{2005ApJ...625..414T}.  A way to achieve this could be
efficient dust growth in the pile-up produced by radiation pressure In
fact, the jump in gas surface density will foster faster and more
efficient dust settling, further enhancing the dust density.  Greatly
increased coagulation rates might be the result.  Large grains
produced in this way could even enter the gap without being visible,
and without being an obstacle to the ionizing radiation needed to
drive the inside-out evacuation of the disk.  However, a much more
detailed model of this scenario is required before any conclusions can
be drawn.

A second possibility is that the inner hole has been made by a planet
orbiting close to the disk rim, and that a pressure bump created by
the planet-disk interaction traps particles in the disk rim
\citep{2006MNRAS.373.1619R}.  However, \citet{2009LPI....40.1477W} has
questioned the effectiveness of this mechanism.

A third possibility would be that torques exerted by one or more
planets located in the inner disk hole in the system transport matter
so quickly to the star, that the average optical depth in the inner
disk remains small even though dust is contained in the accreting
matter \citep{2011ApJ...729...47Z,2011ApJ...727....2P}.

Our model contains a number of simplifications, for example the use of
geometrical cross sections for the dust grains.  We also did not
account for the optical depth of the gas inward of the dust rim.  The
gas is expected to have much lower optical depth than the dust, but at
high enough accretion rates even the gas can become optically thick to
stellar radiation \citep{Muzerolle:2004p1707}.  This effect is rather
weak at the low accretion rates we are concerned with (according to
Muzerolle et al.~it plays virtually no role at $\dot M\simeq
10^{-9}M_{\odot}$/yr), and therefore it is not problematic that we
ignore it.  As with other assumptions in this paper, we have always
made choices that would make it easier to hold back the grains by
radiation pressure.  The conclusion that dust grains cannot be held
back by radiation pressure is therefore a firm one.

\begin{acknowledgements} We would like to thank the anonymous referee
  and E. Chiang for comments on the first version of this manuscript.
  CPD acknowledges financial support from the Max Planck Gesellschaft
  under the SNWG program.  CD would like to thank the Dutch Top
  Research School NOVA (network 2) for financial support and the Leids
  Kerkhoven-Bosscha Fonds for travel support.
\end{acknowledgements}

\end{document}